\documentclass[sigconf=true,review=false,natbib=true,anonymous=false]{acmart}
\usepackage{pifont}
\usepackage{multirow}%
\usepackage{tabularx}
\usepackage{enumitem}
\usepackage{tikz}
\usepackage{fontawesome}
\usepackage{mwe}
\usepackage{subfig}
\usepackage{graphicx}
\usepackage{balance} 
	
\usepackage{soul}
\definecolor{loguipurple}{rgb}{0.274, 0.314, 0.612}
\newcommand{\cmark}{\ding{51}}%
\newcommand{\xmark}{\ding{55}}%

\newcommand{\rqOne}{\textbf{RQ1}}
\newcommand{\rqTwo}{\textbf{RQ2}}
\newcommand{\rqThree}{\textbf{RQ3}}
\newcommand{\rqFour}{\textbf{RQ4}}

\newcommand{\oOne}{\textbf{O1}}
\newcommand{\oTwo}{\textbf{O2}}
\newcommand{\oThree}{\textbf{O3}}
\newcommand{\oFour}{\textbf{O4}}

\newcommand{\oFive}{\textbf{O5}}
\newcommand{\oSix}{\textbf{O6}}
\newcommand{\oSeven}{\textbf{O7}}
\newcommand{\oEight}{\textbf{O8}}

\newcommand{\HG}{\textbf{HG}}
\newcommand{\HL}{\textbf{HL}}
\newcommand{\SG}{\textbf{SG}}
\newcommand{\SL}{\textbf{SL}}
\newcommand{\logui}{\color{loguipurple}\textbf{\texttt{Log}}\texttt{UI}\color{black}}

\newcommand{\remember}{\textit{Remember}}
\newcommand{\analyze}{\textit{Analyse}}
\newcommand{\understand}{\textit{Understand}}
\newcommand{\navigation}{\textit{Navigation}}

\newcolumntype{D}[1]{>{\centering\arraybackslash}p{#1}}
\newcolumntype{L}[1]{>{\arraybackslash}p{#1}}

\definecolor{paperred}{RGB}{137, 23, 31}
\definecolor{papergreen}{RGB}{55, 142, 49}

\newcommand*\circled[1]{\tikz[baseline=(char.base)]{\node[shape=circle,fill=black,inner sep=1pt] (char) {\textcolor{white}{#1}};}}

\newcommand*\circledr[1]{\tikz[baseline=(char.base)]{\node[shape=circle,fill=paperred,inner sep=1pt] (char) {\textcolor{white}{#1}};}}

\newcommand*\circledg[1]{\tikz[baseline=(char.base)]{\node[shape=circle,fill=papergreen,inner sep=1pt] (char) {\textcolor{white}{#1}};}}

\AtBeginDocument{%
  \providecommand\BibTeX{{%
    \normalfont B\kern-0.5em{\scshape i\kern-0.25em b}\kern-0.8em\TeX}}}
    
\copyrightyear{2022}
\acmYear{2022}
\setcopyright{rightsretained}
\acmConference[SIGIR '22]{Proceedings of the 45th International ACM
SIGIR Conference on Research and Development in Information
Retrieval}{July 11--15, 2022}{Madrid, Spain}
\acmBooktitle{Proceedings of the 45th International ACM SIGIR
Conference on Research and Development in Information Retrieval (SIGIR
'22), July 11--15, 2022, Madrid,
Spain}\acmDOI{10.1145/3477495.3531719}
\acmISBN{978-1-4503-8732-3/22/07}


\usepackage{etoolbox}
\makeatletter
\patchcmd{\maketitle}{\@copyrightpermission}{
\begin{minipage}{0.3\columnwidth}
    \href{https://creativecommons.org/licenses/by/4.0/}{\includegraphics[width=0.90\textwidth]{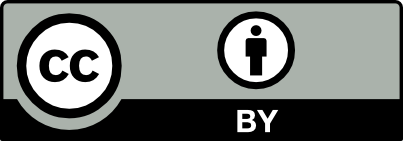}}
  \end{minipage}\hfill
 \begin{minipage}{0.7\columnwidth}
\href{https://creativecommons.org/licenses/by/4.0/}{This work is licensed under a Creative Commons Attribution International 4.0 License.}
  \end{minipage}
}{}{}

\begin{document}
\fancyhead{}
    \thanks{This research has been supported by \textit{NWO VIDI} project \textit{SearchX} (639.022.722) and \textit{NWO} project \textit{Aspasia} (015.013.027).}

\title{Users and Contemporary SERPs: A (Re-)Investigation}
\subtitle{Examining User Interactions and Experiences}

\author{Nirmal Roy}
\affiliation{
    \institution{Delft University of Technology}
    \city{Delft}
    \country{The Netherlands}
}
\email{n.roy@tudelft.nl}
\author{David Maxwell}
\affiliation{
    \institution{Delft University of Technology}
    \city{Delft}
    \country{The Netherlands}
}
\email{d.m.maxwell@tudelft.nl}
\author{Claudia Hauff}
\affiliation{
    \institution{Delft University of Technology}
    \city{Delft}
    \country{The Netherlands}
}
\email{c.hauff@tudelft.nl}

\begin{abstract}
The \textit{Search Engine Results Page (SERP)} has evolved significantly over the last two decades, moving away from the simple \emph{ten blue links} paradigm to considerably more complex presentations that contain results from multiple verticals and granularities of textual information. Prior works have investigated how user interactions on the SERP are influenced by the presence or absence of \emph{heterogeneous} content (e.g., images, videos, or news content), the layout of the SERP (\emph{list} vs. \emph{grid} layout), and \emph{task complexity}. In this paper, we reproduce the user studies conducted in prior works---specifically those of~\citet{arguello2012task} and~\citet{siu2014first}---to explore to what extent the findings from research conducted five to ten years ago still hold today as the average web user has become accustomed to SERPs with ever-increasing presentational complexity. To this end, we designed and ran a user study with four different SERP interfaces: \textit{(i)}~\emph{a heterogeneous grid}; \textit{(ii)}~\emph{a heterogeneous list}; \textit{(iii)}~\emph{a simple grid}; and \textit{(iv)}~\emph{a simple list}. We collected the interactions of $41$ study participants over $12$ search tasks for our analyses. We observed that SERP types and task complexity affect user interactions with search results. We also find evidence to support most (6 out of 8) observations from~\cite{arguello2012task,siu2014first} indicating that user interactions with different interfaces and to solve tasks of different complexity have remained mostly similar over time.
\end{abstract}



\maketitle

\section{Introduction}\label{sec:intro}

    The \textit{Search Engine Results Page (SERP)} has evolved significantly over the last two decades, moving away from the \emph{ten blue links} paradigm, to considerably more complex presentations that contain results from multiple verticals and multiple granularities of textual information (snippets, direct answers, entity cards, etc.)---all interleaved within one page. The incorporation of heterogeneous content in a SERP has been shown to change how users interact with web results~\cite{dumais2010individual,arguello2011methodology,navalpakkam2013measurement,arguello2014effects,liu2015influence,bota2016playing,wu2020providing,salimzadeh2021impact}. How (and where) content is displayed in a SERP affects user interactions as well~\cite{sushmita2010factors,arguello2012task,shao2022linear}. While contemporary web SERPs maintain the original idea of a \emph{list} of items that are ranked in decreasing order of relevance, alternative presentations such as a \emph{grid} layout---as also recently (again) popularised by \href{https://www.you.com}{\textit{You.com}}---have also been explored~\cite{kammerer2014role,siu2014first,xie2017investigating,namoun2018three}.

In addition, past research~\cite{sushmita2010factors,arguello2012task,siu2014first,wu2020providing} has shown that user behaviour on the SERP does not only depend on the \emph{presentation} of information, but also on the \emph{search task} at hand. For a navigational task such as \emph{`find and access the homepage of SIGIR 2022'}, a user---in the ideal case---requires a single query and a single click. Contrast this to an informational task, such as \emph{`good restaurants near the venue of SIGIR 2022'}. This requires the scanning of multiple results, and likely results in further query reformulations to learn more about specific suggestions.

As commercial web search engine SERPs have evolved over time (and thus end users have become accustomed to different types of SERPs), we explore in this paper to what extent user study findings from $5-10$ years ago still hold today. Specifically, we focus our attention on reproducing the experimental setup of two prior studies:~\citet{arguello2012task} (published in 2012) as well as~\citet{siu2014first} (published in 2014)---these both investigated how user interactions
on the SERP are influenced by the presence or absence of heterogeneous content, the layout of the SERP (list vs. grid), and task complexity. Inspired by the two papers we reproduce, our study is guided by the following research questions.

\begin{enumerate}
\item[\rqOne{}] How does a user's interactions with a SERP differ when results are presented in a list and grid layout?
\item[\rqTwo{}] How does task complexity affect user interactions with a SERP?
\item[\rqThree{}] What is the interplay between task complexity and SERP layout on user interactions?
\item[\rqFour{}] How do users perceive the different SERP layouts?
\end{enumerate}

To this end, we conducted a user study with $n=41$ participants that each were given $12$ search tasks of varying complexity (ranging from search tasks of type \remember{} to \analyze{}) to solve with one of four different SERP interfaces: \textit{(i)}~\emph{heterogeneous grid}; \textit{(ii)}~\emph{heterogeneous list}; \textit{(iii)}~\emph{simple grid}; and \textit{(iv)}~\emph{simple list}. We explore whether the following eight observations from~\cite{siu2014first,arguello2012task} about users and their interactions with list vs. grid layouts---and heterogeneous vs. simple results---across different task complexities hold today.

\begin{enumerate}
       \item[\oOne{}] Users fixated significantly more on the grid layout SERP compared to the list layout SERP for completing more complex tasks~\cite{siu2014first}.
    
    \item[\oTwo{}] On the grid layout SERP, users fixated on search results significantly more for completing more complex tasks compared to simple tasks. A similar observation was found for the list layout SERP~\cite{siu2014first}.
    
    \item[\oThree{}] On the list layout SERP, users fixated significantly longer for completing more complex tasks compared to simple tasks. For the grid SERP, there were no significant differences in fixation duration between varying task complexities~\cite{siu2014first}.

     \item[\oFour{}] In the list layout SERP, more complex tasks required significantly greater levels of search interaction: longer search sessions, more clicks on SERP, and more web pages visited~\cite{arguello2012task}.
     
      \item[\oFive{}] In a SERP where web results are arranged in a list layout, users clicked on significantly more vertical results when they were present on the main page of the SERP (blended, heterogeneous display) compared to when they were only present as tabs (non-blended, simple display)~\cite{arguello2012task}. 
      
      \item[\oSix{}] Task complexity did not have a significant effect on user interaction with vertical results in the list layout SERP~\cite{arguello2012task}.
    
    \item[\oSeven{}]  The interplay between task complexity and display of verticals (blended, heterogeneous display vs. non-blended, simple display) did not have a significant effect on user interaction with vertical results in the list layout SERP~\cite{arguello2012task}.
      
      \item[\oEight{}] Neither study~\cite{siu2014first, arguello2012task} found significant differences in user evaluation of the different SERP types, list vs grid layout for the former and blended vs non-blended display for the latter, in their experiments.
     \end{enumerate} 
    
In our user study, we observed that SERP types and task complexity affect user interactions with search results. We also find evidence to support most---6 out of 8---observations from~\cite{arguello2012task,siu2014first}.

\section{Related Work}

\subsection{Task Complexity and User Interactions}
A number of works have focused on the effect of task types on user interactions on SERPs.  
~\citet{buscher2012large} performed a large-scale analysis using query logs to understand how individual and task differences might affect search behaviour. Their findings show that there are cohorts of users who examine search results in a similar way. They also showed that the type of task has a pronounced impact on how users engage with the SERP.~\citet{arguello2012task} observed that the more complex the task, the more users would interact with various components on the SERP whereas~\citet{thomas2013users} found that users tended to examine the result list deeper and more quickly when facing complex tasks.
\citet{jiang2014searching} compared user interactions in relatively long search sessions ($10$ minutes; about $5$ queries) for search tasks of four different types.~\citet{wu2020providing} also observed differences in user interactions with the SERP based on whether they had to look for answers to a factoid question or a non-factoid question. In these studies, the SERPs were composed of web search results in the \textit{de facto} list format. 

\subsection{SERP Presentation and User Interactions}
\citet{sushmita2010factors} observed that positioning (top, middle, bottom) of different verticals on a SERP affects clickthrough rates of users when the verticals (news, image and video) were presented in a blended manner with the web search results.~\citet{arguello2012task} also looked into how task complexity affects user interactions and usage of aggregated vertical results when they are interleaved with web results, versus when they are presented as tabs. On a similar note, they observed that for more complex tasks, users clicked on more vertical results when they were interleaved with web search results.~\citet{bota2016playing} conducted a crowdsourced online user study to investigate the effects of entity cards given ambiguous search topics. They found that the presence of entity cards has a strong effect on both the way users interact with search results and their perceived task workload.
Furthermore,~\citet{levi2018selective} performed a comprehensive analysis of the presentation of results from seven different verticals (including a community question answering vertical) based on the logs of a commercial web search engine. They observed that the community question answering vertical receives on average the highest number of clicks compared to other verticals. \citet{wu2020providing} studied how the presence of answer modules on SERPs affect user behaviour and whether that varies with question types (factoid vs. non-factoid). They found that the answer module helps users complete search tasks more quickly, and reduces user effort. In the presence of answer modules, users' clicks on web search results were significantly reduced while answering factoid questions. 
\citet{shao2022linear} conducted a user study to understand how user interaction is affected by the presence of results in the right rail of a heterogeneous SERP in addition to the traditional web results in the left-rail. They found that users have more interactions with the SERPs, appear to struggle more, and feel less satisfied if they examine the right-rail results.
Overall, findings observed that the presence of verticals and other heterogeneous modalities of results and their position on the SERP affect user interactions. In these studies, results were also presented in the \textit{de facto} list format.

 \citet{kammerer2014role} observed that when web search results are presented in a grid layout, the impact of search result positioning on selecting trustworthy sources is drastically reduced in comparison to the more traditional list approach. Users typically follow a top-down approach when scanning lists, and are more susceptible to select untrustworthy sources if they appear high up in the list. This effect is reduced for a grid-based presentation. However, the authors do not compare different types of tasks, nor do they explore user behaviours when results from various vertical features of the search engines are present on the SERP.~\citet{siu2014first} compared the eye-tracking data of grid and list SERP layouts with two types of tasks (informational vs. navigational), and investigated potential differences in gaze patterns. The \textit{`F-shaped'} pattern was less prominent on the grid in comparison to the list layout. These two studies explore how user interactions change when web results are presented in a grid vs. a list layout. They do not, however, include vertical results in their study. 

\section{Methodology}
To address our four overarching research questions as outlined in~\S\ref{sec:intro}, we conducted a user study with $n=41$ participants. Each participant was assigned to one of four experimental search interface conditions \textbf{(interfaces: \textit{between-subjects})}, and completed 12 search tasks \textbf{(tasks: \textit{within-subjects})}.

\begin{figure*}[t!]
\vspace*{-2.5mm}
\hspace*{-4.1mm}
\includegraphics[width=1.04\textwidth]{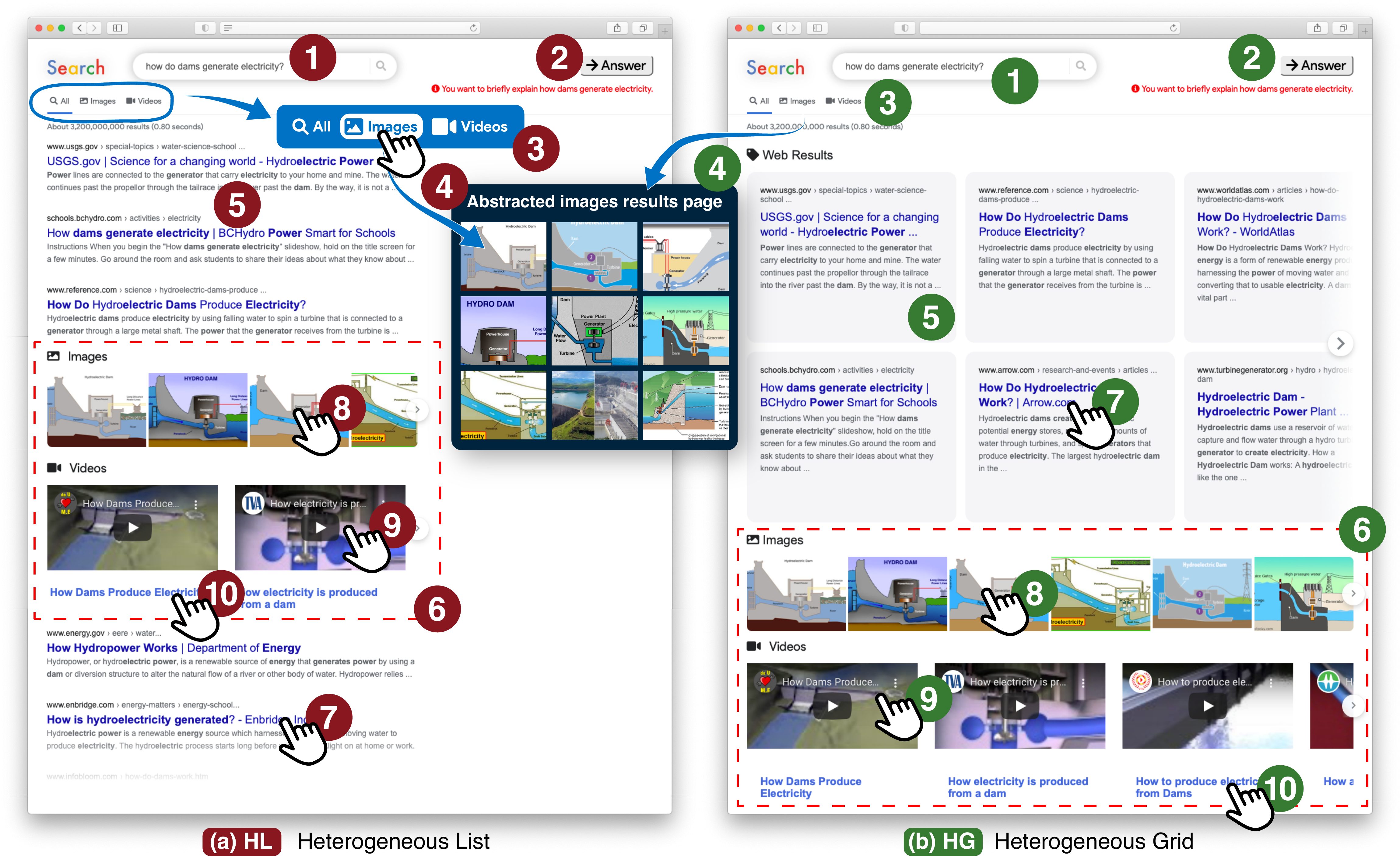}
\caption{Examples of both the \textit{(a)} list-based and \textit{(b)} grid-based interfaces trialled. Note the inclusion of links for the separate \faSearch~\textit{All} (as shown), \faImage~\textit{Images}, and \faVideoCamera~\textit{Videos} result pages. Heterogeneous content is displayed in red boxes, and is \textit{not} present in the two homogeneous content interface conditions (\SG{} and~\SL{}). Circled numbers correspond to the narrative of Section~\ref{sec:system}.} 
\label{fig:interface}
\end{figure*}

Our four experimental search interface conditions considered the \textbf{\textit{layout type}} (\textit{list-} vs. \textit{grid-}based layout) and the \textbf{verticals present} on the SERP (\textit{heterogeneous content} vs. \textit{homogeneous content}). These combinations result in the interface conditions outlined below, with examples of the two layout types presented in Figure~\ref{fig:interface}, with further details provided in \S\ref{sec:system}.

\begin{enumerate}[leftmargin=0.70cm]

    \item[\SL{}]{\textbf{Simple List} Considered as our baseline interface condition (the standard and widely used \textit{ten blue links}~\cite{hearst2009_search}), this interface presents results in a list, with each result presented one under the other. All results are \textit{web results}, and as such are \textit{homogeneous} in terms of presented content.}
    
    \item[\SG]{\textbf{Simple Grid} The same homogeneous approach to content is taken as for \SL{}, but with results presented in a \textit{grid-based} approach. Instead of scrolling along the vertical, participants subjected to this interface scroll along the \textit{horizontal}.}
    
    \item[\HL{}]{\textbf{Heterogeneous List} Similar in approach to \SL{}, \HL{} presents results in a list. However, different verticals are mixed in with the standard web results. Beyond web results, heterogeneous content used in this study includes \textit{image} and \textit{video} results.}
    
    \item[\HG{}]{\textbf{Heterogeneous Grid} Similar to \HL{} but now the content is displayed in grid form, with \textit{web-based} results appearing in a grid, before additional image and video content.}
    
\end{enumerate}

\subsection{Search Interface Design and System}\label{sec:system}
Given the above, our goal is to find out to what extent the observations from prior studies by both~\citet{arguello2012task} and~\citet{siu2014first} are valid after almost a decade of SERP design evolutions (and additions). To operationalise our four experimental interface conditions, we first needed to create a SERP template design that closely mirrors the design of a contemporary web search engine.

For this study, we selected the \textit{Google} SERP as it presents information recognisably, and commands approximately 92\% market share.\footnote{\url{https://gs.statcounter.com/search-engine-market-share}. All URLs in this paper were last checked on 2022-02-14.} A replica template was created with particular attention paid to the colour schemes, fonts, width and height of components---as well as the spacing between them. The end result was a highly realistic template of a contemporary SERP, on which we based all of our study's results pages.\footnote{Templates are released for future user studies, available at \url{https://github.com/roynirmal/sigir2022-serp-reproducibility}.}

\vspace*{1mm}\noindent\textbf{SERP Template Overview} Figure~\ref{fig:interface} presents the SERP template used, presenting results for the query \texttt{`how do dams generate electricity?'}. Present are examples for both list- \textit{(Figure~\ref{fig:interface}\textit{(a)})} and grid-based \textit{(Figure~\ref{fig:interface}\textit{(b)})} interfaces.

A query box is provided \circled{1}. However, this is disabled and provides no functionality for this study. It does however display the query terms that were used to derive the presented results \textit{a priori} (see \S\ref{sec:method:infoneeds}). This is presented next to the \textit{information need} for a given task \circled{2}, alongside which there is a button to take the participant to the next stage of the experiment. The SERP template also provides links to additional results pages, namely \faImage~\textit{Images} and \faVideoCamera~\textit{Videos} \circled{3}, emulating the setup of the study by~\citet{arguello2012task}. As shown at \circled{4}, a grid-based layout is shown for both image and video pages, as is the norm in commercial web search engines such as Google and \textit{Bing}. For images, a total of $16$ were displayed (in a $4\times4$ grid); for videos, a total of nine were shown (in a $3\times3$ grid).

On the SERP, standard web results are presented \circled{5}, with the \textit{de facto} $10$ \textit{results per page (RPP)} provided. For list-based interfaces \circledr{5} (\SL{} and \HL{}), web results are displayed in the standard way, with one result following the other down the left rail of the SERP. Grid-based interfaces \circledg{5} (\SG{} and \HG{}) present the results in a \textit{carousel} user interface component (emulating the setup of~\citet{siu2014first}), where the $10$ results are arranged in the form of a $5\times2$ grid. A total of six ($3\times2$) results were visible \textit{above-the-(vertical)-fold}; access to the remaining four results ($2\times2$) was made available through use of a button to scroll across.

As denoted by the red dashed boxes in Figure~\ref{fig:interface} \circled{6}, heterogeneous content is also added to the SERP template. Present only in experimental search interface conditions \HL{} and \HG{}, these components provided inline image and video results to the participants. 
Like web results in the grid-based interfaces, these were also scrollable, mimicking the behaviour of contemporary web search engine SERPs. Sufficient content was placed within these components to ensure two complete scrolls could be completed; the number of images displayed varied as their widths were variable. On interface condition \SG{}, image and video components were placed under the \textit{third} web result; on interface condition \HG{}, they were placed directly underneath the web results grid.

The SERP template was also fully interactive---participants could click on links of web results \circled{7}, with a new browser tab then opening to present the page at the linked URL. In addition, images and videos within the SERP could also be interacted with. Clicking on an image \circled{8} took the participant to the webpage containing the image (again, in a new tab). Videos, all sourced from \textit{YouTube}, could be played on the SERP itself \circled{9}, with the necessary infrastructure in place to enable such functionality. If the participant wished to view the video on the YouTube website itself, they could click the link underneath \circled{10} to do so. Again, YouTube links opened a new tab in the participant's browser.

\vspace*{1mm}\noindent\textbf{SERP Definition} Note that for each query, there are three unique results pages to replicate the study of~\citet{arguello2012task}. These are the results \textit{`landing'}, or \faSearch~\textit{All} page, containing the web results (and additional components, for interface conditions \HL{} and \HG{})---as shown in Figure~\ref{fig:interface}, as well as the \faImage~\textit{Images} and \faVideoCamera~\textit{Videos} pages. From hereon in, we refer to a SERP as the `landing' page, containing web results. To replicate the study of~\citet{siu2014first}, the `landing' page itself is sufficient. 

\vspace*{1mm}\noindent\textbf{Capturing Interactions and Experiences} Integrated with the SERP template was \logui{}~\cite{maxwell2021logui}, a framework-agnostic JavaScript library for capturing different interactions and other events within a web-based environment. \logui{} was configured to capture a series of mouse events (including hovers, clicks, and scrolls) over the various components of the SERP. Interactions on components included (but were not limited to) web results, images and video contents (including the capturing of the playback, pausing, and completion of YouTube videos). We also recorded interactions to (and on) the supplementary \faImage~\textit{Images} and \faVideoCamera~\textit{Videos} pages. Browser-wide events were also captured, and included the ability for us to compute the time spent \textit{away} from the SERP---when participants would click on a document/image/video link, which would open a new tab.

Experience data was captured via a number of \textit{Qualtrics}\footnote{\url{https://www.qualtrics.com/}} surveys; a pre- and post-experiment survey were completed by each participant, in addition to the small post-task summary the participants had to write. More details on these questions and the flow of the experiment can be found in \S\ref{sec:method:procedure}. Our setup ensured that participants would jump between the Qualtrics surveys and SERPs as and when required.

\vspace*{2mm}\noindent\textbf{Static SERPs} As alluded to with the disabled query box \circled{1}, our experimental setup featured no programmable backend or search functionality. This meant that there was \textit{no} additional querying functionality. We served manually curated SERPs that we produced \textit{a priori} for each of the $12$ search tasks we asked participants to undertake. This setup ensured that all study participants viewed the same results (a setup also chosen in prior studies, such as those by~\citet{sushmita2010factors} and~\citet{wu2020providing}). While making the search experience somewhat less realistic, it did provide us with the benefit of not having to deal with participants submitting diverse queries. The design also removed a confounding variable and allowed us to address our four RQs by calculating the user interaction measures on a fixed set of web results, images, and videos.

\vspace*{-2mm}
\subsection{Search Tasks}\label{sec:method:infoneeds}
For this study, we used four different types of information need: a \textbf{\textit{Navigational}} type \textit{(where individuals seek to find particular websites)}, and three different \emph{informational} categories, belonging to the \textbf{\remember{}} \textit{(involving the retrieval, recognition, and recalling of relevant knowledge)}, \textbf{\understand{}} \textit{(constructing meaning from information sources)}, and \textbf{\analyze{}} \textit{(involving the breakdown of information info constituent parts, and determining how they related to one another)} categories. For each category, we produced three unique information needs. This led to a total of $12$ information needs which are listed in Table~\ref{tab:info}. Particular attention was paid to designing tasks that enticed participants to not only look at web results but also at image and video search results as well.

The choice of our information needs is based on the study designs used by both~\citet{arguello2012task} and~\citet{siu2014first}. More specifically,~\citet{arguello2012task} designed a series of tasks that required different levels of diversity of \textit{information} to complete---as well as different amounts of search effort. Tasks were grounded in the revised \textit{Bloom taxonomy}, as outlined by~\citet{anderson2001taxonomy}.\footnote{The Bloom taxonomy is typically used to create educational materials.} Search tasks were informational and belong to the \remember{}, \understand{}, and \analyze{} categories. In addition,~\citet{siu2014first} employed just two categories of tasks for their study---navigational and informational. Our design thus combines the setups of both prior works that we wish to examine.


\begin{table*}[!t]
    \centering
    \caption{Overview of information needs and their type. The rightmost column shows the most popular query obtained from our query selection pilot study, outlined in Section~\ref{sec:pilot}. Numbers in parentheses indicate how many crowdworkers (\boldmath{$n=25$}) submitted the most popular query variation. Here, \textit{Nav.}=\textit{Navigation}, \textit{Remem.}=\textit{Remember}, and \textit{Underst.}=\textit{Understand}.}
    \vspace*{-2mm}
    \label{tab:info}
    
    \renewcommand{\arraystretch}{1.2}
    \footnotesize
    \begin{tabular}{L{0.5cm}L{10.2cm}L{6cm}}
        \toprule
        \textbf{Type} &
        \textbf{Information Need} & \textbf{Most Popular Query Variation} \\
        \midrule\midrule
        \multirow{3}{*}{\hspace*{2mm}\rotatebox{90}{\textbf{\textit{Nav.}}}} & You want to find the homepage of Andrew Zimmern, the chef. & \texttt{andrew zimmern chef} (14) \\
         & You want to find the page of Air Jordan on the Nike website. & \texttt{nike air jordan} (14) \\ 
         &  You want to find the page displaying the Flixbus route map in Europe. & \texttt{flixbus europe route map} (6)\\ \hline
        
        \multirow{3}{*}{\vspace{-1mm}\hspace*{2mm}\rotatebox{90}{\textbf{Remem.}}} & You want to know where is the pituitary gland located in the body. & \texttt{where is the pituitary gland} (9) \\
        & You want to find out what clothes the famous cartoon character Mickey Mouse typically wears. & \texttt{what clothes does mickey mouse wear} (5) \\
        & You want to find out how to calculate the volume of an ellipsoid. & \texttt{ellipsoid volume formula} (4) \\ \hline
        \multirow{3}{*}{\vspace*{-1mm}\hspace*{2mm}\rotatebox{90}{\textbf{\textit{Underst.}}}} & You want to find out the steps required to make a paper airplane. &  \texttt{how to make a paper airplane} (10) \\
         &  You want to briefly explain how dams generate electricity. & \texttt{how do dams generate electricity} (17)\\
         & You want to find out how to prevent shower mirrors from fogging. & \texttt{stop shower mirror fogging} (3) \\ \hline
        \multirow{3}{*}{\vspace*{-10mm}\hspace*{2mm}\rotatebox{90}{\textbf{\analyze{}}}} & You want to get into martial arts, but you have no fighting experience. Which form of martial arts is more suitable for beginners? & \texttt{best martial arts for beginners} (3)  \\
        & You want to find out the main things to look for while installing solar panels on the roof of a house. & \texttt{things to consider before installing rooftop solar panels} (1)   \\ 
        & You want to buy a new camera lens for taking professional pictures of your friend. Which camera lenses are best for portrait photography? & \texttt{best camera lenses for portrait photography} (3) \\

        \bottomrule
    \end{tabular}
    \vspace*{-3mm}
\end{table*}


\subsection{Query Selection and SERP Curation}\label{sec:pilot}
To ensure that participants received helpful search results, we required a common search query for each. To this end, we ran a small crowdsourced pilot study on the \textit{Prolific} platform\footnote{\url{https://www.prolific.co/}}. This pilot had $n=25$ workers, with the design largely inspired by the study reported by~\citet{bailey2016uqv100}. Workers took approximately $10$ minutes to complete the task and were paid at the hourly rate of GBP$8.00$ for their time. All $12$ information needs were presented to the workers. They were instructed to type the query terms that they would issue to their web search engine of choice if they were seeking information to address the information need. Collected queries were then normalised (case normalisation, stripped punctuation, whitespace cleanup) and passed through the \textit{Bing Spell Check API} to generate a final canonical form of each submitted query. Subsequently, we determined the most frequently occurring query variation for the $12$ tasks, taking this query forward as the one to use for the next stage of our study. These are listed in the parentheses in Table~\ref{tab:info}.\footnote{All query variations belonging to \textit{solar panels installation} information need of \analyze{} tasks were slightly different from each other, as crowdworkers tended to submit natural language queries for this information need. We manually picked the query that we deemed to be the best one for this particular information need.}



\vspace*{2mm}\noindent\textbf{Curating SERPs} We then used a combination of the \textit{Bing Web Search API}, \textit{Bing Image Search API}, and \textit{Bing Video Search API} to curate a collection of: \textit{web} (title, snippet text, and target URL); \textit{image} (source image and document URL); and \textit{video} (video source URL) results for each of the $12$ queries. Snippet text was truncated to the equivalent of two sentences/lines, as this has been previously shown to be a good trade-off in terms of providing a sufficient \textit{information scent} and encouraging interaction (i.e., clicks)~\cite{maxwell2017snippets}. Video links were filtered to YouTube only, as utilising only one video content provider reduced complexity for playback on our SERPs. Any URLs that proved non-functional or redirected to a \texttt{404} page were also removed. The content was then placed on our SERP templates, allowing us to construct SERPs, an image results page, and a video results page for each query. SERP variations for all four search interface conditions were produced.


\vspace*{-3.3mm}
\subsection{Experimental Procedure}\label{sec:method:procedure}
The $12$ search tasks undertaken by each participant were preceded and followed by pre- and post-experiment surveys. We first performed screen and browser viewport resolution checks, requiring that all participants use a maximised browser window with a resolution of $1920\times1080$ or greater. This ensured that we could guarantee the SERPs displayed to the participants could be viewed without scrollbars along the horizontal. If the checks were successful, participants began the experiment by providing basic demographic information and were also asked minor questions on their search engine usage, specifically on what components on a contemporary SERP they often make use of. In addition, we asked what their preferred search engine is. They were then randomly assigned to one of the four search interface conditions (\SL{}, \SG{}, \HL{}, or \HG{}).

Participants were primed to summarise their findings after each search task (in no more than $50$ words). Upon acceptance of this instruction, the first search task began, with a SERP similar to the one presented in Figure~\ref{fig:interface}\textit{(a)} or Figure~\ref{fig:interface}\textit{(b)}. With the selected query \circled{1} and information need \circled{2} present, participants then began to examine the content. Participants were \emph{not} given a minimum or maximum amount of time to search. We reiterate that they were also \emph{not} given the opportunity to issue their queries. Once they were satisfied with what they had found, they clicked the \texttt{\faArrowRight Answer} button at the top of the SERP \circled{2}, and entered their summary. Once complete, the next task began. This process was repeated for the remaining $11$ tasks which were displayed to them in random order. Other researchers have also employed randomisation for condition allocation to minimise topic ordering effects~\cite{lagun2014towards,wu2020providing}.

\begin{table*}[t!]
    \centering
    
    \caption{Results of a factorial mixed ANOVA, where interface is between-subjects, and task is within-subjects variable. A \cmark{} indicates significant effect (\boldmath{$p<0.05)$} on the particular user interaction and \xmark{} indicates no significant effect.}
    \vspace*{-2mm}
    \label{tab:anova}
    
    \renewcommand{\arraystretch}{1.1}
    \footnotesize
    \begin{tabular}{L{5.6cm}D{3.6cm}D{3.6cm}D{3.6cm}}
        \toprule
        \textbf{User Interactions} &
        \textbf{SERP Main Effect} &
        \textbf{Task Main Effect} & \textbf{B/W SERP \& Task} \\
        \midrule\midrule
 
 Web results clicks & \cmark{} (F $=4.27, p=0.01)$  & \cmark{} (F $=4.18, p=0.01)$  & \xmark{}\\
 Mean web result reading time (s) & \xmark{} & \cmark{} (F $=3.97, p=0.01)$ & \xmark{}
 \\
 Mean session duration (s) &\xmark{} & \cmark{} (F $=12.72, p<0.0001)$ & \xmark{} \\

 Mean web result hover duration (s) & \xmark{} & \xmark{} & \xmark{} \\ \hline
  Image clicks (SERP)& \xmark{} & \cmark{} (F $=7.24, p = 0.004)$ & \xmark{}
 \\
  Video clicks (SERP) & \xmark{} & \xmark{} & \xmark{} \\
  Image hovers (SERP) & \xmark{} & \cmark{} (F $=6.98, p = 0.009)$& \xmark{} \\
  Video hovers (SERP) &\xmark{} & \xmark{} & \xmark{}
 \\ \hline
   Image clicks (image results page) &\xmark{} & \cmark{} (F $=4.66, p = 0.01)$ & \xmark{} \\
 Video clicks (video results page) & \xmark{} & \xmark{} & \xmark{} \\
  Image hovers (image results page) & \xmark{} & \cmark{} (F $=5.39, p = 0.01)$ & \xmark{}  \\
  Video hovers (video results page) & \cmark{} (F $=3.36, p = 0.02)$ & \xmark{} & \xmark{}    \\
 
        \bottomrule
    \end{tabular}
    \vspace*{-3mm}
\end{table*}

After the search tasks had been completed, participants then moved on to the post-experiment survey. We used the sub-scales from O'Brien's \textit{Engagement Scale}~\cite{o2010development,o2010influence} as was done by ~\citet{arguello2012task}. These are aimed at eliciting their evaluation of the interface they used on the following aspects of engagement: \emph{focused attention}; \emph{perceived usability}; \emph{experience}; \emph{aesthetics}; and \emph{felt involvement}. The engagement scale was originally designed to evaluate shopping websites, and hence we modified/removed the statements pertaining to shopping to suit our needs. For example, we changed the original statement (belonging to \emph{aesthetics} sub-scale) \textit{``This shopping website was aesthetically appealing''} to \textit{``The layout of the results page is aesthetically appealing''}. For all statements in the sub-scales, participants indicated their level of agreement ($1$=\textit{strongly agree}; $5$=\textit{strongly agree}). We also used the \emph{search effectiveness} sub-scale used by~\citet{arguello2012task} to evaluate how effective the interfaces were in helping participants find information. In total, we used $26$ statements from the six sub-scales to elicit user evaluation of the search interfaces. The reliability scores (Cronbach’s Alpha) for the sub-scales are reported in Table~\ref{tab:serp}. They were also asked to rate the perceived usefulness of web, image and video results.




\vspace*{-2.5mm}
\subsection{Study Participants}
Like our pilot, we recruited participants from the Prolific platform. Our $n=41$ participants were native English speakers from the United Kingdom, with a $95\%$ approval rate on the platform, and had a minimum of $250$ prior successful task submissions. From our participants, $32.5\%$ identified as female, and $67.5\%$ as male. The mean age of our participants was $36.5\pm9.7$, with a minimum age of $22$ and a maximum of $68$. $92\%$ of participants listed \textit{Google} as their preferred search engine, with the remaining $8\%$ identified as \textit{DuckDuckGo} users. With respect to the highest completed education level, $51.2\%$ possess a Bachelors (or equivalent), $24.4\%$ have a Masters (or equivalent), $19.5\%$ have a high school degree, and 4.9\% have an Associate (or equivalent). $95\%$ of participants cited using web results on a contemporary SERP, $78\%$ made use of image results, with $37\%$ citing that they used video results.

In our random assignment, $11$ participants were assigned to \HG{}, with ten participants each assigned to \HL{}, \SL{}, and \SG{}. The experiment lasted on average $40$ minutes for the $41$ participants. Like our pilot participants, they were compensated at the rate of GBP $8.00$ per hour. All participants who registered completed the study; post-hoc checks confirmed that they had provided sensible answers for each task, and as such we approved all who took part for payment. As such, our base analyses are reported over $41*12=492$ search sessions and their corresponding interaction logs.
\section{Results and discussion}
For our analyses\footnote{All data and code pertaining to our analyses are available \textcolor{blue}{\href{https://drive.google.com/drive/folders/1dtK6T1fnxFeLiQbUx9CAl8TJTn3gKqxw?usp=sharing}{here}}.}, we conduct a series of mixed factorial ANOVA tests to observe if task complexity, SERP types or the interplay between them have a significant effect on interactions. We follow up the ANOVA with post-hoc $t$-tests with Bonferroni correction ($p<0.05$) to observe where significant differences occur. We evaluate if observations~\oFour{}-\oSeven{} also hold in grid SERPs, \HG{} and \SG{}.

\vspace*{-3mm}
\subsection{\rqOne{}: SERP Type and User Interactions}
\label{sec:serp_interact}

Table~\ref{tab:anova} presents results that are relevant to our first three research questions. Here, a \cmark{} indicates a significant effect ($p<0.05)$ on the particular user interaction, and a \xmark{} indicates no significant effect.

As seen in Tables~\ref{tab:anova} and ~\ref{tab:serp}, different SERP types do \emph{not} have a significant effect on user interactions except for: \textit{(i)} the number of web results clicked (row \textbf{I}, Table~\ref{tab:serp}); and \textit{(ii)} the number of hovers on videos present in the video results page (\textbf{XV}, Table~\ref{tab:serp}). Post-hoc tests reveal that participants in the \HL{} condition have significantly more web result clicks than their \HG{} and \SG{} counterparts (\textbf{I}, Table~\ref{tab:serp}). \HL{} participants also have longer web result reading times compared to participants in any of the other SERP conditions (\textbf{II}, Table~\ref{tab:serp})---albeit not significant. Furthermore, participants with the list interfaces (\HL{} and \SL{}) have a greater number of hovers over web results compared to \HG{} and \SG{}. As a result, we cannot confirm \oOne{} where~\citet{siu2014first} found significantly more fixation counts on the grid interface than on the list interface. We note that, since we did not record eye gaze data, we are approximating fixation counts by user interactions such as web result clicks and snippet text hovers, as mouse position has been shown to correlate with gaze positions in prior studies~\cite{rodden2007exploring, rodden2008eye,navalpakkam2013measurement}. One of the possible reasons for the difference in observation with \oOne{} can be that our participants are more familiar with the standard list layout of web results, as a majority use Google as their main search engine.

\citet{arguello2012task} do not compare user interactions with web results on heterogeneous SERPs vs. simple SERPs. However, we observe that participants using the simple SERP interfaces (\SG{} and \SL{}) scan web search results to lower depths than those of their heterogeneous interface counterparts (\textbf{IV}, Table~\ref{tab:serp}). The lack of information (i.e., fewer verticals) on the SERP requires participants to scan web results to a greater depth in the ranked list.
     
Based on Table~\ref{tab:serp} (\textbf{VII-X}), we find that on average \HL{} participants interact more with image and video results that are present on the SERP compared to their \HG{} counterparts. \SG{} and \SL{} participants interact more with vertical results present in the image and video results page than those of their heterogeneous counterparts (\textit{XI-XIV}, Table~\ref{tab:serp}). Post-hoc tests also reveal that \SL{} participants have significantly more hovers on video results on the video results page than participants in the other SERP conditions (\textbf{XIV}, Table~\ref{tab:serp}). The lack of vertical results on the SERP makes the participants interact with them in the respective vertical results pages which shows that our informational needs indeed require participants to seek out image and video search results as well. \HG{} and \HL{} participants seem to be satisfied with vertical results present on the SERP and the former barely interacted with vertical results present in the respective results pages (\textit{XI-XIV}, Table~\ref{tab:serp}). Looking at overall interactions with vertical results (adding interactions with vertical results present on the SERP and the vertical results pages for \HG{} and \HL{}), we see that \HG{} and \HL{} have slightly more interactions than \SG{} and \SL{} respectively. This difference is not significant, but we do see a trend in the line of \oFive{} where~\citet{arguello2012task} observed a higher number of vertical result clicks when they were blended with the web results in the SERP. This is compared to when they were only present on the respective vertical results page. On a side note, the higher interactions with vertical results present on the SERP by \HL{} participants compared to \HG{} participants (\textbf{VII-X}, Table~\ref{tab:serp}, also depicted in Figure~\ref{fig:interact-DMAX}\textit{(c)}) can be attributed to the fact that images and videos in \HL{} SERPs appear in the middle of the web results (between rank $3$ and $4$) whereas they appear below the web results in \HG{} SERPs. Participants in the latter interface expend comparatively more effort to access the vertical results, thereby reducing their interaction. We leave further analysis on the effect of positioning of vertical results on user interaction for future work.

Addressing \rqOne{}, we found that the interface has a significant main effect on the clicks on web results and hovers on videos on the video results---page but not on other user interactions. 
 
\vspace*{-2.5mm}
\subsection{\rqTwo{}: Task Complexity and User Interactions}
\label{sec:task_interact}

Table~\ref{tab:task} shows that the informatioon needs of the \textit{Analyse} type, which are the most complex among our information needs, warrant most web result clicks (\textbf{I}), web result dwell time (\textbf{II}) and session duration from participants (\textbf{III}). Table~\ref{tab:anova} shows that the main effect of task complexity on these interactions is significant. Participants reach greater web result click depth (\textbf{IV}, Table~\ref{tab:task}) for \textit{Analyse} tasks, albeit not significant. Post-hoc tests reveal that \textit{(i)} \textit{Analyse} tasks receive significantly more web result clicks than \remember{} tasks; \textit{(ii)} \analyze{} and \understand{} tasks lead to significantly higher web result dwell times than \textit{Navigation} tasks; and \textit{(iii)} the session duration for \analyze{} and \understand{} tasks are significantly higher than for \textit{Navigational} tasks, while the session duration for \analyze{} tasks is also significantly greater than that for \remember{} tasks. Overall, we find that user interactions on web search results increase as the complexity of information needs increase which is inline with the observations of ~\citet{arguello2012task} and we can partially confirm \oFour{}.

~\citet{arguello2012task} did not include \textit{Navigational} tasks in their experiments. We argue that they can be considered as tasks requiring the lowest level of cognition, and as such follow the trend of \oFour{}---they receive the least interaction among all task categories. The only exception to this was web result clicks---the nature of the task requires participants to click web result links to ascertain that they found the correct page.

We approximate fixation duration in \oThree{} by observing hover duration over the web results, akin to fixation count in~\S\ref{sec:serp_interact}.
Although participants hover longer over web results (\textbf{V}, Table~\ref{tab:task}) and snippet text (\textbf{VI}) for \remember{} tasks compared to other tasks, the difference across tasks is not significant. Moreover, the mean hover duration on web results (snippet and title) for participants belonging to the grid SERP types (\HG{} and \SG{}) is longer than for those belonging to the list SERP types (\textbf{VI}, Table~\ref{tab:serp}). As seen from Table~\ref{tab:anova}, the interplay between SERP type and task complexity do not have a significant effect on hover duration over web results. As a result, we can only partially confirm \oThree{} where~\citet{siu2014first} also do not find significant differences in fixation duration for grid layout for the tasks but they \textit{did} find significantly longer fixation duration on the list layout for more complex tasks.

\begin{table*}[!t]
    \centering
    
    \caption{User interactions for different interfaces across all tasks. $\dagger$ indicates that there is a significant main effect of SERP layout on that particular user interaction. $^\mathcal{HG}$, $^\mathcal{HL}$, $^\mathcal{SG}$, $^\mathcal{SL}$ indicate significant difference with \textbf{HG}, \textbf{HL}, \textbf{SG} and \textbf{SL} respectively. Maximum values for each interaction is highlighted in \textbf{bold}. }
    \vspace*{-2mm}
    \label{tab:serp}
    
    \renewcommand{\arraystretch}{1.1}
    \footnotesize
    \begin{tabular}{L{0.5cm}L{4.6cm}D{2.65cm}D{2.65cm}D{2.65cm}D{2.65cm}}
        \toprule
        
        & & \multicolumn{4}{c}{\textbf{Interface Condition}} \\
        
        \textbf{Row} &
        \textbf{Interaction} &
        \textbf{HG} &
        \textbf{HL} & \textbf{SG} &
        \textbf{SL} \\
        \midrule\midrule
 
\textbf{I}& Web result clicks$\dagger$ & $11.27 (\pm 8.43)^\mathcal{HL}$ & $\mathbf{21.30 (\pm 8.99)}^{\mathcal{HG},\mathcal{SG}}$ & $18.20 (\pm 9.91)^\mathcal{LH}$ & $18.70 (\pm 11.89)$ \\  
\textbf{II}& Mean web result reading time (s) & $17.96 (\pm 12.74)$ & $\mathbf{27.00 (\pm 28.82)}$ & $16.82 (\pm 9.05)$ & $25.09 (\pm 10.64)$ \\ 
\textbf{III}& Mean session duration (s) & $94.89 (\pm 56.43)$ & $106.96 (\pm 46.85)$ & $98.35 (\pm 47.52)$ & $\mathbf{109.24 (\pm 64.55)}$ \\ 

\textbf{IV}&Maximum web result click depth & $3.36 (\pm 2.60)$ & $3.62 (\pm 1.45)$ & $\mathbf{4.22 (\pm 1.89)}$ & $4.15 (\pm 1.61)$ \\ 
\textbf{V} & Total web result hovers & $66.55 (\pm 29.77)$ & $83.40 (\pm 61.38)$ & $122.70 (\pm 87.07 )$ & $\mathbf{124.40 (\pm 100.79)}$ \\
\textbf{VI}&Mean web result hover duration (s) & $\mathbf{2.91 (\pm 6.95)}$ & $2.49 (\pm 4.74)$ & $2.20 (\pm 4.96)$ & $0.80 (\pm 0.69)$ \\  

\hline
 \textbf{VII}& Image clicks (SERP) & $0.82 (\pm 1.17)$ & $\mathbf{2.10 (\pm 2.38)}$   & - & - \\
 \textbf{VIII}& Video clicks (SERP) & $1.55 (\pm 2.81)$ & $\mathbf{11.20 (\pm 34.04)}$ & - & - \\
 \textbf{IX}&  Image hovers (SERP) &  $13.27 (\pm 14.88)$ & $\mathbf{16.30 (\pm 16.73)}$ & - & - \\
 \textbf{X}& Video hovers (SERP) & $6.18 (\pm 9.66)$ & $\mathbf{13.40 (\pm 22.62)}$ & - & - \\ \hline
\textbf{XI}& Image clicks (image results page) & $0.00 (\pm 0.00)$ & $0.40 (\pm 0.70)$ & $0.70 (\pm 1.06)$ & $\mathbf{1.10 (\pm 1.45)}$ \\ 
\textbf{XII}& Video clicks (video results page) & $0.00 (\pm 0.00)$ & $0.00 (\pm 0.00)$ & $0.00 (\pm 0.00)$ & $\mathbf{0.70 (\pm 1.49)}$ \\ 
\textbf{XIII}& Image hovers (image results page) & $0.00 (\pm 0.00)$ & $4.10 (\pm 10.67)$ & $8.80 (\pm 18.58)$ & $\mathbf{13.00 (\pm 19.96)}$ \\ 
\textbf{XIV}& Video hovers (video results page)$\dagger$ & $0.00 (\pm 0.00)^\mathcal{SL}$ & $0.00 (\pm 0.00)^\mathcal{SL}$ & $0.10 (\pm 0.32)^\mathcal{SL}$ & $\mathbf{2.80 (\pm 4.85)}^{\mathcal{HG},\mathcal{HL},\mathcal{SG}}$ \\ \hline
\textbf{XV} &Usefulness of image results & $2.45 (\pm 0.93)$ & $2.60 (\pm 0.84)$ & $2.40 (\pm 0.84)$ & $\mathbf{2.90 (\pm 1.37)}$ \\ 
\textbf{XVI} &Usefulness of video results & $2.27 (\pm 1.10)$ & $2.30 (\pm 1.06)$ & $\mathbf{2.40 (\pm 0.84)}$ & $2.10 (\pm 0.74)$ \\ 
\textbf{XVII} &Usefulness of web results & $4.64 (\pm 0.50)$ & $\mathbf{4.70 (\pm 0.67)}$ & $4.50 (\pm 0.71)$ & $\mathbf{4.70 (\pm 0.48)}$ \\ 
\hline
\textbf{XVIII} & Focused attention ($\alpha = 0.846) $ & $3.45 (\pm 0.80)$ & $\mathbf{4.33 (\pm 0.49)}$ & $3.70 (\pm 1.20)$ & $3.40 (\pm 0.86)$ \\ 
\textbf{XIX} & Experience ($\alpha = 0.791) $  & $\mathbf{4.39 (\pm 0.39)}$ & $4.12 (\pm 0.78)$ & $3.98 (\pm 0.49)$ & $4.15 (\pm 0.68)$ \\ 
\textbf{XX} & Aesthetics  ($\alpha = 0.942) $ & $3.45 (\pm 0.77)$ & $\mathbf{3.62 (\pm 1.13)}$ & $3.33 (\pm 0.91)$ & $3.50 (\pm 0.66)$ \\ 
\textbf{XXI}& Felt involved  ($\alpha = 0.647) $& $\mathbf{4.00 (\pm 0.56)}$ & $\mathbf{4.00 (\pm 0.67)}$ & $3.80 (\pm 0.74)$ & $3.77 (\pm 0.55)$ \\ 
\textbf{XXII}& Effectiveness  ($\alpha = 0.735) $& $\mathbf{4.35 (\pm 0.40)}$ & $4.30 (\pm 0.36)$ & $3.92 (\pm 0.43)$ & $4.12 (\pm 0.61)$ \\ 
\textbf{XXIII} & Usability  ($\alpha = 0.891) $& $3.18 (\pm 0.38)$ & $2.68 (\pm 1.28)$ & $2.87 (\pm 0.66)$ & $\mathbf{3.22 (\pm 0.69)}$ \\

        \bottomrule
    \end{tabular}
\end{table*}

\begin{table*}[]
    \centering
    
    \caption{User interactions for different task complexity across all search interfaces. $\dagger$ indicates that there is a significant main effect of task complexity on that particular user interaction. $^\mathcal{N}$, $^\mathcal{R}$, $^\mathcal{U}$,
$^\mathcal{A}$ indicate significant difference with navigational, remember, understand and analyse tasks. Maximum values for each interaction is highlighted in \textbf{bold}. }
    \vspace*{-2mm}
    \label{tab:task}
    
    \renewcommand{\arraystretch}{1.1}
    \footnotesize
    \begin{tabular}{L{0.5cm}L{4.5cm}D{2.65cm}D{2.65cm}D{2.65cm}D{2.65cm}}
        \toprule
        \textbf{Row} & 
        \textbf{Interactions} &
        \textbf{Navigational} &
        \textbf{Remember} & \textbf{Understand} &
        \textbf{Analyze} \\
        \midrule\midrule
 
\textbf{I} & Web result clicks $\dagger$ & $4.00 (\pm 2.65)$ & $3.76 (\pm 3.25)^\mathcal{A}$ & $4.12 (\pm 2.55)$ & $\mathbf{5.34 (\pm 4.07)}^\mathcal{R}$ \\ 
 \textbf{II} & Mean web result reading time (s)$\dagger$ & $14.99 (\pm 11.62)^{\mathcal{A},\mathcal{U}}$ & $20.57 (\pm 21.18)$ & $24.05 (\pm 22.64)^\mathcal{N}$ & $\mathbf{26.91 (\pm 29.18)}^\mathcal{N}$ \\ 
\textbf{III} & Mean session duration (s)$\dagger$ & $69.95 (\pm 52.66)^{\mathcal{A},\mathcal{U}}$ & $97.11 (\pm 76.00)^\mathcal{A}$ & $106.90 (\pm 62.22)^\mathcal{N}$ & $\mathbf{134.75 (\pm 73.98)}^{\mathcal{N},\mathcal{R}}$ \\ 
\textbf{IV} & Maximum web result click depth  & $3.32 (\pm 2.59)$ & $3.88 (\pm 2.27)$ & $3.85 (\pm 2.37)$ & $\mathbf{4.27 (\pm 2.77)}$ \\ 
\textbf{V} & Mean web result hover duration (s) & $0.86 (\pm 1.98)$ & $\mathbf{3.26 (\pm 14.43)}$ & $2.05 (\pm 8.28)$ & $2.31 (\pm 9.89)$ \\ 
\textbf{VI} & Mean snip. text hover duration (s) & $0.08 (\pm 0.22)$ & $\mathbf{0.12 (\pm 0.24)}$ & $0.07 (\pm 0.11)$ & $0.08 (\pm 0.15)$ \\ 

\hline
\textbf{VII} & Image clicks (SERP)$\dagger$ & $0.17 (\pm 0.38)^\mathcal{A}$ & $\mathbf{0.44 (\pm 1.00)}^\mathcal{A}$ & $0.12 (\pm 0.40)$ & $0.00 (\pm 0.00)^{\mathcal{N},\mathcal{R}}$ \\ 
\textbf{VIII} &  Video clicks (SERP) & $0.00 (\pm 0.00)$ & $1.15 (\pm 5.59)$ & $0.12 (\pm 0.64)$ & $\mathbf{1.88 (\pm 11.40)}$ \\ 
\textbf{IX} & Image hovers (SERP)$\dagger$ & $1.10 (\pm 2.45)^\mathcal{R}$ & $\mathbf{4.83 (\pm 10.20)}^{\mathcal{N},\mathcal{U},\mathcal{A}}$ & $1.07 (\pm 2.53)^\mathcal{R}$ & $0.54 (\pm 1.98)^\mathcal{R}$ \\ 
\textbf{X} &  Video hovers (SERP) & $0.88 (\pm 2.61)$ & $\mathbf{4.49 (\pm 15.71)}$ & $1.90 (\pm 5.51)$ & $1.29 (\pm 4.47)$
 \\ \hline
\textbf{XI} & Image clicks (image results page) $\dagger$ & $0.07 (\pm 0.35)$ & $\mathbf{0.32 (\pm 0.65)}^{\mathcal{U},\mathcal{A}}$ & $0.12 (\pm 0.40)^\mathcal{R}$ & $0.02 (\pm 0.16)^\mathcal{R}$ \\ 
\textbf{XII} & Video clicks (video results page) & $0.00 (\pm 0.00)$ & $0.00 (\pm 0.00)$ & $\mathbf{0.15 (\pm 0.69)}$ & $0.02 (\pm 0.16)$ \\ 
\textbf{XIII} & Image hovers (image results page)$\dagger$ & $1.15 (\pm 4.11)$ & $\mathbf{4.37 (\pm 10.92)}^\mathcal{A}$ & $0.73 (\pm 2.55)$ & $0.07 (\pm 0.35)^\mathcal{R}$ \\ 
\textbf{XIV} & Video hovers (video results page) & $0.02 (\pm 0.16)$ & $0.00 (\pm 0.00)$ & $\mathbf{0.56 (\pm 2.21)}$ & $0.12 (\pm 0.64)$ \\

        \bottomrule
    \end{tabular}
\end{table*}

Among interactions with vertical results present on the SERP (\textbf{VII-X}, Table~\ref{tab:task}), we observe that \remember{} tasks receive the most interactions on average. Post-hoc tests reveal that: \textit{(i)} participants click significantly more on images present on the SERP (\textbf{VIII}, Table~\ref{tab:task}) for \remember{} and \textit{Navigational} tasks compared to \analyze{}; and \textit{(ii)} they hover significantly more on images present on the SERP (\textbf{X}, Table~\ref{tab:task}) for \remember{} tasks compared to all other task categories. 
For images present on image results page, we again observe significantly more image clicks (\textbf{XII}, Table~\ref{tab:task}) and hovers (\textbf{XIV}, Table~\ref{tab:task}) for \remember{} tasks compared to the more complex \understand{} or \analyze{} tasks. Findings regarding user interactions with vertical results (present on the SERP and the vertical results pages) and their relationship with task complexity is contrary to the observations of~\citet{arguello2012task}, and hence we cannot confirm~\oSix{}. The high interaction with vertical results for \remember{} tasks together with the fact that participants hover over web results and snippet text longer (on average) for the same task (\textbf{V} \& \textbf{VI}, Table~\ref{tab:task}) shows that participants prefer to address information needs of the \remember{} type by either hovering over web results and interacting with verticals rather than clicking the link.~\citet{arguello2012task} do not observe hover duration in their analysis.

To answer \rqTwo{}, we find that task complexity does have a significant effect on several user interactions. With participants interacting more with web results as tasks get more complex, we observe significantly more interactions with image results for \remember{} tasks compared to more complex \analyze{}/\understand{} tasks.



\begin{figure*}[!t]
    \includegraphics[width=\linewidth]{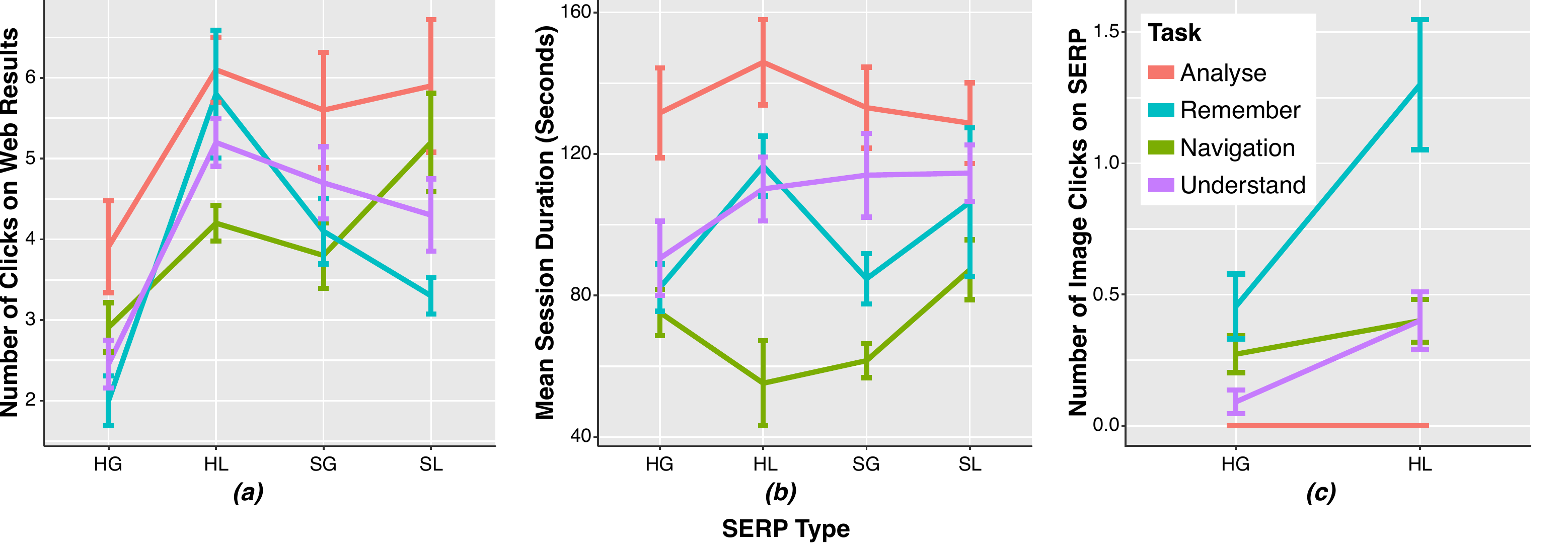}
    \vspace*{-5mm}
    \caption[Caption for LOF]{Interaction plots, showing effects of SERP types and task complexity over: \textit{(a)} clicking on web results; \textit{(b)} the mean session duration (in seconds); and \textit{(c)} clicks on images presented on the SERP.\protect\footnotemark}
    \label{fig:interact-DMAX}
\end{figure*}
\footnotetext{Figure ~\ref{fig:interact-DMAX} in the \href{https://dl.acm.org/doi/10.1145/3477495.3531719}{SIGIR proceedings} version has a mistake---the legends of \textit{Remember} and \textit{Navigation} are flipped. This is the correct version of the plot.}

\vspace*{-2.5mm}
\subsection{\rqThree{}: Task Complexity, SERP Type and User Interactions}

As seen in Table~\ref{tab:anova}, we do not observe a significant effect of the interplay between SERP types and task complexity on user interaction with web results or verticals which is similar to what~\citet{arguello2012task} found. Hence, we can confirm \oSeven{}.
     
From Figure~\ref{fig:interact-DMAX}\textit{(a)}, we observe that participants across all SERP types click the most web results for \analyze{} tasks (in line with \oFour{}). For each task type, \HG{} participants click the least number of web results and for most tasks participants with grid SERPs click lower ranked web results than those with list SERPs (in contrary to \oOne{}). Approximating fixation count by web result clicks, as done in~\S\ref{sec:task_interact}, we see for each SERP type, the complex \analyze{} tasks receive more interaction than the less complex \remember{} or \understand{} tasks. Although pairwise comparisons do not show a significant difference in web result clicks between different tasks for each SERP, we observe a trend similar to \oTwo{}---more complex tasks requiring higher document clicks. From Figure~\ref{fig:interact-DMAX}\textit{(b)}, we see that participants across all SERP types take longest to finish \analyze{} tasks and least amount of time to finish \textit{Navigational} tasks (also in line with \oFour{}{}). Finally, Figure~\ref{fig:interact-DMAX}\textit{(c)} corroborates our findings from~\S\ref{sec:task_interact}, as we see that participants across both SERP types interact most with image results for \remember{} tasks compared to other tasks. As mentioned earlier, this observation is contrary to what~\citet{arguello2012task} observed in their study (\oSix{}).

\begin{figure}[!t]
    \includegraphics[width=\linewidth]{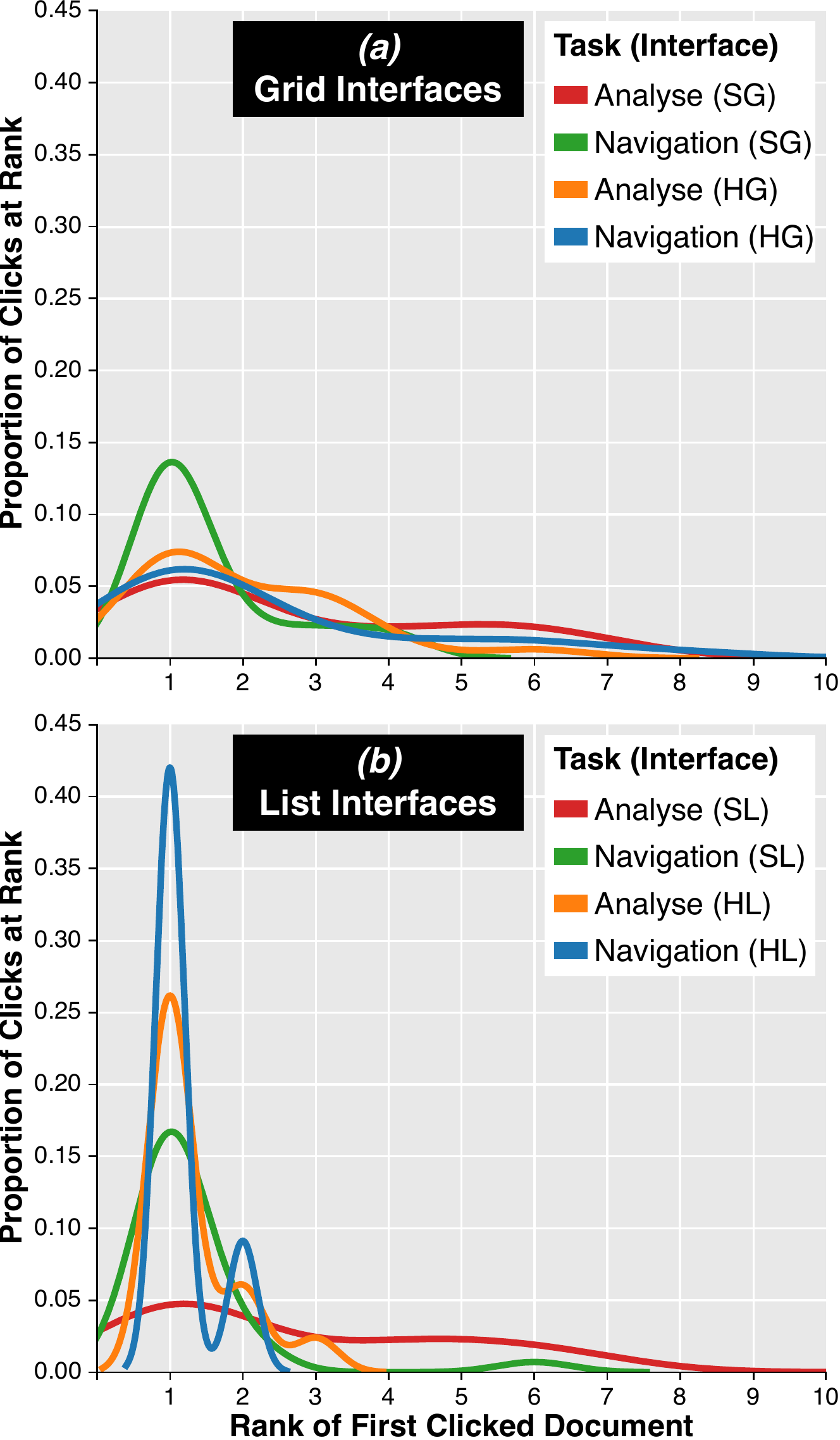}
    \vspace*{-5mm}
    \caption{Distribution of ranks of the first clicked web results for participants over both grid-based interfaces SL and HL \textit{(a)}, and list-based interfaces SG and HG \textit{(b)}.}
    \label{fig:first_click-DMAX}
\end{figure}

In Figure~\ref{fig:first_click-DMAX}, we plot the distribution of where (which rank) participants made their first click of web results for \navigation{} and \analyze{} tasks. The \HL{} SERP is the only one where the web results are ``broken'' by vertical results at rank three, and as a result, we observe that most of the first clicks for both tasks appear before rank four (subplot \textbf{(b)} of Figure~\ref{fig:first_click-DMAX}). For the other SERPs, the first click distribution for the tasks is more uniform. This is especially prominent for \analyze{} tasks, where we see due to the absence of verticals on \SL{} SERP (comparing \textit{Analyse HL} and \textit{Analyse SL} in subplot \textbf{(b)} of Figure~\ref{fig:first_click-DMAX}) participants are willing to go further down the list before their first click. We also expect a peak around the first result for \textit{Navigation} tasks, which is true for all SERP types except \textit{Navigation HG} in subplot \textbf{(a)}. Either the participants using that SERP type prefer to not click a lot as is evident from Table~\ref{tab:serp} (fewest web result clicks by \HG{} participants), or they chose to explore more before their first click. It has been observed in earlier works~\cite{joachims2017accurately,kammerer2014role} that participants have a trust bias for list SERPs (they click on web results appearing higher up the ranked order). The trust bias had been found previously to be less prevalent in grid SERPs~\cite{kammerer2014role}. We also find evidence of similar user interaction in subplot \textbf{(b)} compared to subplot \textbf{(a)} where participants are more open to exploration before their first click. To conclude, we find for \rqThree{}, that the interplay between SERP types and task complexity does not have a significant effect on user interactions.
    





\subsection{\rqFour{}: Perceived Experience of SERPs}

Turning our attention to the post-experiment surveys, we observe little
difference in participant ratings of the systems (\textbf{XVIII-XXIII}, Table~\ref{tab:serp}). This is in line with both~\citet{arguello2012task} and \citet{siu2014first}, who also did not find significant differences in user ratings for different interfaces. Therefore, we can confirm \oEight{}. 

We also observe that web search results on average are perceived to be more useful (\textbf{XVII}, Table~\ref{tab:serp}) than image or video results (\textbf{XV-XVI}, Table~\ref{tab:serp}). This is in line with the click behaviour of participants. Across all SERP types, they clicked on more web results than they did on images or videos.~\citet{arguello2012task} also found the overall number of vertical clicks to be lower than that on web results. Image results were perceived to be more useful by \SL{} participants followed by their \HL{} counterparts (\textbf{XV}, Table~\ref{tab:serp}), which is reflected in their behaviour as well. While the former has the most interactions with images present on the image results page (\textbf{XI-XIII}, Table~\ref{tab:serp}) compared to participants in other cohorts, the latter interacted most with images present on the SERP (\textbf{VII-IX}, Table~\ref{tab:serp}).

    
    
    



\section{Conclusion}

\noindent\textbf{Summary} In this work, we set out to answer the question of how four different types of SERP and four different types of tasks of varying levels of complexity affect user interaction with web, image and video results. We also explore whether observations about users and their interactions from the studies of~\citet{arguello2012task} and \citet{siu2014first} hold with contemporary SERPs. We observed the following findings with respect to our research questions.

\begin{enumerate}
    \item[\rqOne{}] The SERP has a significant main effect on the number of clicks on web results and the number of hovers on videos on the video results page, but not on other user interactions.
    
    \item[\rqTwo{}] Task complexity has a significant effect on user interactions. While participants interact more with web results as the task becomes more complex, we observe significantly more interactions with image results for \remember{} tasks compared to the more complex \analyze{} or \understand{} tasks.
       
    \item[\rqThree{}] The interplay between SERP types and task complexity does not have a significant effect on user interactions.
    
    \item[\rqFour{}] There is little difference in the evaluation of the four SERP types by participants.
\end{enumerate}

Out of eight observations, we found evidence to confirm two (\oSeven{}, \oEight{}), with partial evidence for a further four (\oTwo{}, \oThree{}, \oFour{}, \oFive{}). These findings indicate that the user interactions over different interfaces for solving tasks of varying complexity have remained mostly similar over time. However, we employed different information needs---and recruited different participants---from the prior studies. Nevertheless, the evidence contrary to \oOne{} and \oSix{} has interesting implications---introducing SERPs that users are not familiar with might result in a decrease in interaction. Although the grid layout can present search results in a condensed format (displaying more items in a given screen space compared to the list layout), users might still end up exploring more in the familiar list layout. Additionally, interactions with vertical results are not only dependent on the complexity of the tasks, but also the type of information need.
As we observed, certain simpler tasks might warrant more interaction with vertical results than more complex tasks~\cite{sushmita2010factors}.

\vspace*{1mm}\noindent\textbf{Reproducing IIR studies} Several variables exist that might affect the observations of an IIR study. An unexhaustive list includes the selection of users, interfaces, and task types. Although both~\citet{arguello2012task} and ~\citet{siu2014first} described how their respective interfaces looked, they did not point to any resources which would help us replicate them. Moreover, we believe that the more users become familiar with a particular interface, the more important it is to present a similar interface to them during a study examining their behaviours. As mentioned in Section~\ref{sec:system}, we have created templates of SERPs that resemble \texttt{google.com} and \texttt{you.com}, and released them for further use. We believe our templates will be useful for the community to eliminate confounding variables in IIR studies that might arise due to SERP presentation. Secondly,~\citet{arguello2012task} and~\citet{siu2014first} did not mention the entire set of tasks used in their studies. As a result, we came up with our tasks of different complexity, as presented in Table~\ref{tab:info}. Two studies by~\citet{urgo2019anderson, urgo2020effects} both list examples of tasks pertaining to different complexities which also offer useful resources for future IIR studies. Our tasks differ with respect to the fact that we designed tasks that specifically enticed participants to not only look at web
results, but also to image and video search results as well. It is important to have a fixed set of tasks and similar interfaces to reproduce and enable reliable comparison of observations (e.g., the number of queries, documents opened, etc.) with prior IIR studies. Lastly, in most cases, it will not be possible to have the same participants while reproducing IIR studies. Crowdsourcing provides a solution for capturing user interactions as it has been shown that there is little difference in the quality between
crowdsourced and lab-based studies~\cite{zuccon2013crowdsourcing}. Power analysis can be used to determine the number of participants required given the experimental conditions of a particular study. It also might be useful to release experimental logs from these studies, after careful ethical checks and considerations. This will permit future researchers to examine them closely, and use them to develop, for example, models of user interaction and search behaviour.

\vspace*{1mm}\noindent\textbf{Limitations and Future Work}
There are several areas with scope for future refinement. First, although we tried to select information needs that cover a broad range of topics, we cannot be certain that the results generalise to information needs with other characteristics. Second, we did not provide querying functionality to users---and hence it will be worthwhile to explore if that has an overall effect on user interactions. Thirdly, the positions of vertical results on the main page of the SERP were fixed, and we know from previous work~\cite{sushmita2010factors,shao2022linear} that user interactions with verticals is affected by where they are displayed on the SERP. In the future, we aim to investigate varying the position of verticals on list and grid interfaces, and their effect on user interaction. The findings from this study can be further applied to designing and evaluating SERP presentations and the placement of heterogeneous content. Understanding and modelling user interactions will also help us work on methodologies for interface optimisation~\cite{wang2013incorporating} and SERP evaluation, along the same veins of prior studies~\cite{chuklin2016incorporating,azzopardi2018measuring,thomas2018better,zhang2020models,sakai2020retrieval}.

\bibliographystyle{ACM-Reference-Format}
\balance
\bibliography{references}

\end{document}